\documentclass{article}
\usepackage{spconf,amsmath,graphicx}
\usepackage{multirow}
\usepackage{siunitx}
\title{Sample drop detection for distant-speech recognition with asynchronous devices distributed in space}
\name{Tina Raissi$^{1 }$ \qquad Santiago Pascual$^{2}$ \qquad Maurizio Omologo$^{3}$}

\address{
	$^{1}$ Human Language Technology and Pattern Recognition, RWTH Aachen University, Aachen, Germany\\ 
	$^{2}$ Universitat Polit\`ecnica de Catalunya, Barcelona, Spain \\
	$^{3}$ Center for Information and Communication Technology, Fondazione Bruno Kessler, Trento, Italy}

\begin{document}
\ninept
\maketitle
\begin{abstract}
In many applications of multi-microphone multi-device processing, the synchronization among different input channels can be affected by the lack of a common clock and isolated drops of samples.\ In this work, we address the issue of sample drop detection in the context of a conversational speech scenario, recorded by a set of microphones distributed in space.\ The goal is to design a neural-based model that given a short window in the time domain, detects whether one or more devices have been subjected to a sample drop event.\ The candidate time windows are selected from a set of large time intervals, possibly including a sample drop, and by using a preprocessing step.\ The latter is based on the application of normalized cross-correlation between signals acquired by different devices.\
The architecture of the neural network relies on a CNN-LSTM encoder, followed by multi-head attention.\ The experiments are conducted using both artificial and real data.\ Our proposed approach obtained F1 score of $88\%$ on an evaluation set extracted from the CHiME-5 corpus.\
A comparable performance was found in a larger set of experiments conducted on a set of 
multi-channel artificial scenes.


\end{abstract}
\section{Introduction}
\label{sec:intro}
 
Distant multi-microphone conversational Automatic Speech Recognition (ASR) in real home environments has attracted a considerable interest in the research community in the past decade.\ There is a large number of ASR systems deployed successfully in applications which require robustness to different sources of degradation of the input signal.\ Moreover, distant-ASR introduces many new complex challenges to solve, mainly due to environmental acoustics and unpredictable noisy conditions, often represented by interfering speakers~\cite{McDonough_book}.\ In order to improve the recognition performance, it is very common to adopt a multi-microphone multi-device setting which allows to observe the scene from different locations in space.\  
For this purpose, the challenges and corpora such as CHiME, DIRHA, and AMI/AMIDA~\cite{barker2018fifth,ravanelli2015dirha,carletta2005ami} were created to address a wide spectrum of research topics related to scenarios such as automatic transcription of speech in office and home environments, with spontaneous speech input, different noisy and reverberant conditions as well as microphone and device placement configurations~\cite{watanabe2017new,li2015robust}.\ In the specific case of the CHiME-5 challenge, we are in the presence of simultaneous recordings of different real conversational scenarios from multiple microphone arrays, distributed in rather large spaces.\ The recording sessions in this case are done by using 6 Kinect microphone arrays and binaural in-ear microphone pairs for each of the 4 speakers.\ The signals recorded by each Kinect microphone array are sample-synchronous within the device.\ However, the different devices can be subjected to asynchrony \cite{Plinge}, due to both small clock speed variations and sample drop events.\ The latter aspect causes misalignment between different signals, creating additional hurdles for tasks such as voice activity detection, beamforming, and any other multi-channel processing that would require synchrony at sample or frame level~\cite{medennikov2018stc, Vincent}.\
It is also worth noting that such misalignment is additional to an intrinsic one that characterizes audio signals acquired by microphones largely spaced one to another, i.e., a shift due to propagation time delays related to the geometry of the problem.\ This can strongly depend on the spatial location of each sound source, which is typically represented by both active speakers and coherent noise sources, and is continuously changing in experimental scenarios such as CHiME-5.
In principle, clock drift, sample drop, and sound source localization~\cite{Cobos} could be jointly addressed with the aim of obtaining a rather accurate re-alignment between all signals.\ Under controlled conditions, e.g., using artificial multi-microphone audio datasets, this joint goal can be addressed, also thanks to fully available ground-truth information.\ On the contrary, in a real context the problem becomes highly challenging, and generally it is not supported by accurate ground-truth information, such as speaker location and drop duration itself.\ Just to give a rough idea about the relevance of the problem, in the case of the CHiME-5 challenge, the shift among different devices due to sample drops exceeds one second in many sessions, while the duration of each drop can range from a few milliseconds to more than hundred milliseconds.\

As discussed in the following, in order to identify temporal intervals that can be affected by sample drop, one can apply standard cross-correlation techniques, which generally need to process rather large context windows (e.g., 10-15 seconds duration), in order both to accurately detect and to eventually quantify the drop duration.\ A crucial aspect is represented by the time accuracy with which the drop is detected.\ For this purpose, the analysis on a shorter context window is strongly required.\ The main focus of this paper is on the sample drop detection within a short context window, which is tackled by adopting a neural classifier.\ The design of the architecture has been conceived to respond to a two-fold problem.\ First, the model has to learn a latent representation of the signal that allows for the comparison between the signal from a device affected by a sample drop event, and the signals coming from the other devices.\ Secondly, the system has to attain robustness to the resulting time shift.\

The paper is structured as follows.\ After a short overview of the relation to the prior works, Section \ref{sec:clTask} describes the binary classification task, cross-correlation based method and the neural solution.\ We then present the dataset, the conducted experiments and the relative results, in Section \ref{sec:exp}, followed by our conclusions.

\section{RELATION TO PRIOR WORK}
\label{sec:prior}
The relevance of the sample drop detection task in applications of multi-device multi-microphone processing is not restricted merely to the research  related to CHiME-5.\ Nowadays, many ASR systems rely on multiple devices distributed in the environment, which transmit eventually the signals to the Cloud.\ Sometimes, these devices are low-cost and characterized by possible interrupts in their operational activity.\ The communication step itself can also be affected by packet loss, for instance in the case of limited internet access.\ Furthermore, there are two different research fields to which the general concept of the detection of missing data can relate.\ However, they both differ from our approach.\ According to the best of the authors knowledge, there is no similar prior work for the sample drop detection in distributed microphone arrays for distant-ASR.\
 
 A first group of studies is done for packet loss concealment algorithms, and the perceptual evaluation of the speech transmission quality in the communication networks.\ The term \textit{discontinuity} defined as one of the three orthogonal speech quality dimensions~\cite{waltermann2013dimension}, describes the isolated distortion introduced by packet loss due to the transmission errors or network delay.\ The predictive model designed with both handcrafted features and deep learning methods estimates the packet loss rate and explains the effect of the transmission errors on the speech quality~\cite{mittag2019quality, mittag2018non}.
 
 Another related field is in the framework of \textit{missing feature techniques}, where it is stated that a robust ASR system must acknowledge that some of the spectro-temporal regions are dominated by noise, and hence incomplete or missing.\ These techniques normally consider a first step for the estimation of a mask which divides the input signal into reliable and unreliable regions~\cite{kim2011mask,seltzer2004bayesian,park2009spatial}.\ In order to deal with the unreliable regions normally the two \textit{marginalization} and \textit{imputation} methods are proposed~\cite{cooke2001robust,li2015robust}.

\begin{figure}[htb]
		\centering 

	\includegraphics[width=0.9\linewidth, trim={0cm 3cm 8cm 0cm},clip]{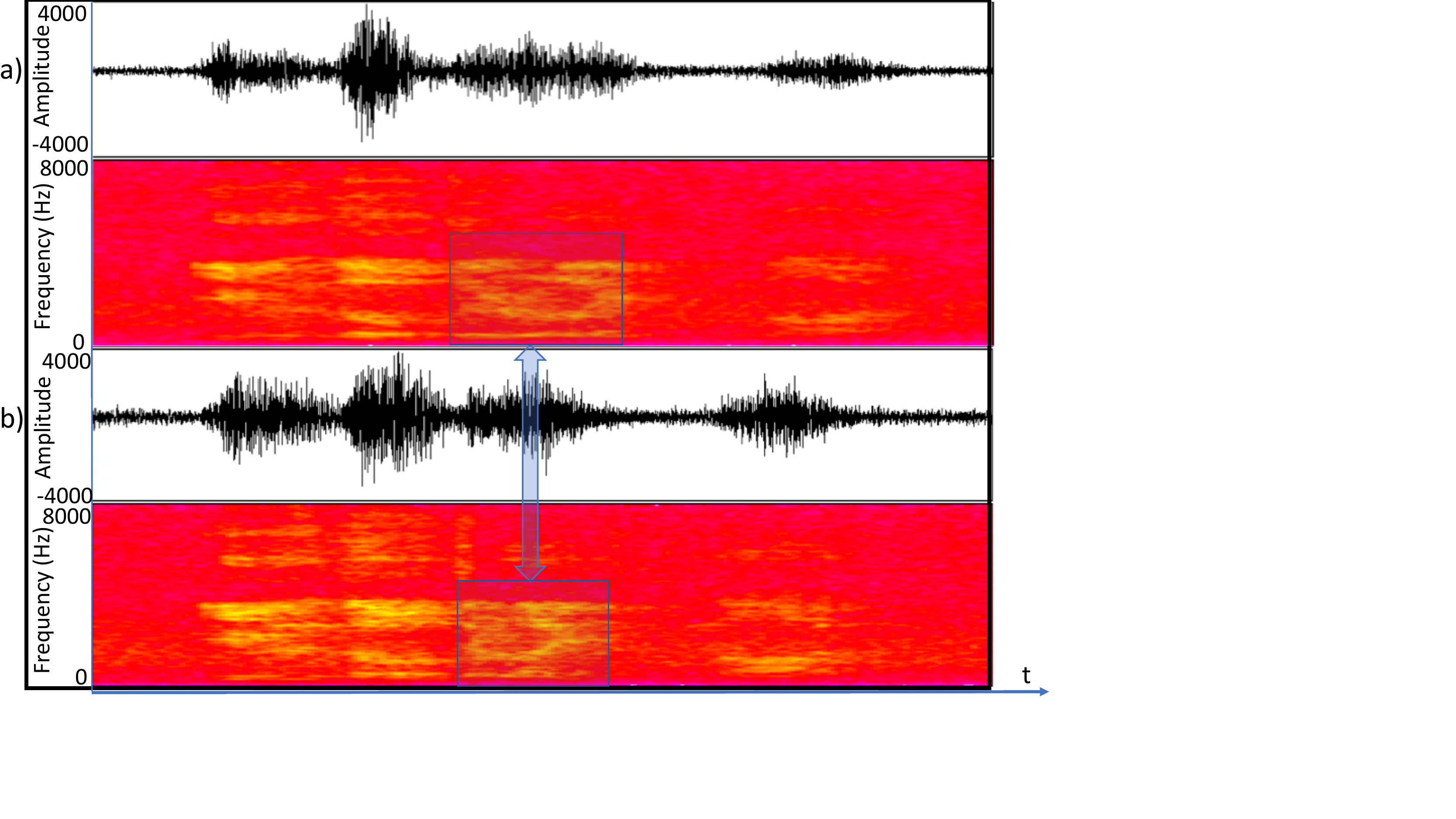}
	    \caption{{\small Signals and related spectrograms referred to recordings acquired by two asynchronous devices for CHiME-5. The signal of case b) is affected by a sample drop, which is highlighted by the shorter vowel sequence in the middle of the sentence (see blue rectangles) and by a different spectrographic alignment after that time instant.}}
		\label{fig:prob_def}
\end{figure}

\vspace{-0.5cm}
\section{Classification Task}
\label{sec:clTask}
 
 
Let $x(t)$ be a source signal, and $y_{m,k}(t)$ the corresponding signal acquired at time sample $t$ by the $m$-th microphone of the device $k$, with $m \in \{1, \cdots, M\}$ and $k \in \{1, \cdots, K\}$, where $M=4$ and $K=6$ in CHiME-5.\ Denote by $Y_{m,k}(\cdot)$ the Short Time Fourier Transform (STFT) of the acquired signal.\ Given a context window of length $\ell$, within a time interval of length $T$, with $\ell \ll T$, our goal is to decide whether a device has been subjected to a sample drop event, as depicted in Figure~\ref{fig:prob_def}.\

We denote by $\text{C}(\cdot, \cdot)$ a function which correlates or compares $|Y_{m,hyp}(\cdot)|$ of a \textit{hypothesis} device with the remaining \textit{reference} devices $|Y_{m,k}(\cdot)|$, for all $k \neq hyp$.\ The binary classification task can be defined as follows:
\begin{equation}
    \label{eq:general}
    \text{Comb}_{k}\left(\mathcal{F} \left( \text{C}\left( \left| Y_{m,hyp}(\cdot)\right|, \left| Y_{m,k}(\cdot\right) \right| \right) \right)  \rightarrow \{0,1\},
\end{equation}
where $\mathcal{F}$ and $\text{Comb}$ are a binary classifier and a combination strategy, e.g. averaging, respectively.\ Furthermore, we assume that the input to the function $\text{C}(\cdot, \cdot)$ is the log-magnitude spectrum $20 \log_{10} \left| Y_{m,k}(\cdot)\right|$, in decibel.\ 

\begin{figure}[t]	
	\centering  
		\includegraphics[width=0.9\linewidth, trim={0cm 0cm 8cm, 3cm},clip]{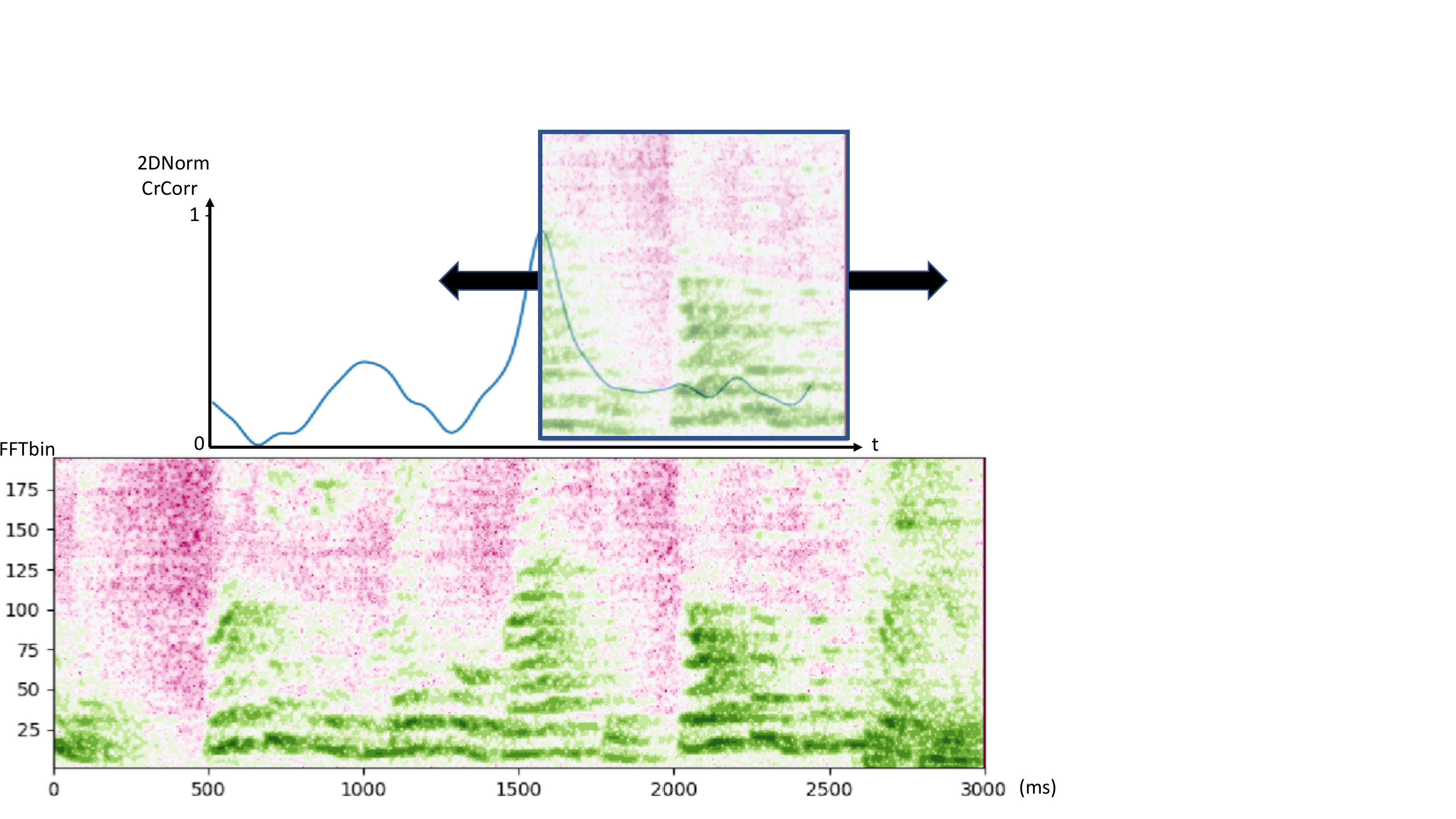}
        \caption{{\small Example of application of 2D normalized cross-correlation to spectrographic patterns extracted from two device microphones. 
        For each time instant, a correlation value is computed between a different time-shifted pattern and a portion of the spectrogram (i.e., lower representation) representing the reference device signal. The resulting correlation function is reported in blue. The time location of its peak highlights the exact shift between the two signals.}}
		\label{fig:cross-correlation}	
\end{figure}


\subsection{Cross-correlation method}
\label{ssec:classic}

The normalized cross-correlation method
adopted in this work, derives from a standard approach
to pattern matching and feature detection often deployed in application fields based on image processing~\cite{gonzalez2008digital,Brunelli,Lewis}.


Given the log-magnitude spectrum $20\log_{10}\left|Y_{m,k}(.) \right|$, it is possible to represent the signal in terms of spectrogram, whose portions are then processed as images, namely to find the best match shift between a pattern (see Figure \ref{fig:cross-correlation}) extracted from a signal of the device $hyp$, and a longer temporal sequence, related to the device $ref$, which is very likely covering the absolute time range of the former one.
Due to different propagation time delays from active sound sources to the microphones of each device, and to minor drift effects due to clock mismatch, in general the two spectrographic representations are not synchronized in time.\ 
Under noisy and reverberant conditions, they can differ substantially, in particular in the case of distant devices.\ Nevertheless, a global similarity is often preserved in terms of temporal relationships among most significant speech contents (e.g., formants and voice onsets), which leads to a higher correlation value corresponding to the correct time shift.
In the case of a sample drop that affects one sequence, this match drastically changes, due to an entire vertical slice of the spectrogram which has been lost in one of the two channels.\ However, this is the key aspect of the proposed method, since a new correlation peak is often found at a time shift that differs from the previous one (observed before the loss) by the exact duration of the loss.
In order to increase the robustness of the method, we 
combine
the analysis outputs obtained for each device by correlating the corresponding signal with each of the other $K-1$ device signals.
As a result, this processing produces $K$ cumulative functions, one per device.

The described method is effective both to identify rather large intervals (e.g., 10-15 seconds) that are likely to include a possible drop, and to provide an estimate of its duration. However, in our preliminary work we observed some limitations when one  aims to derive a more accurate location in time (e.g., less than one second resolution) of this drop. For this reason, in this work we explore the application of this normalized cross-correlation method for a preprocessing before the application of a neural classifier whose goal is to eventually decide between loss and no-loss, in a short context window of a few seconds. 

 

\subsection{Neural-based method}
\label{ssec:neuro}
As mentioned in the last Section, the intervention of the neural model is carried out over a short context window, once the cross-correlation method has identified a set of candidate large time intervals.\   

\begin{figure}[t]
    \centering
    \includegraphics[width=0.9\linewidth]{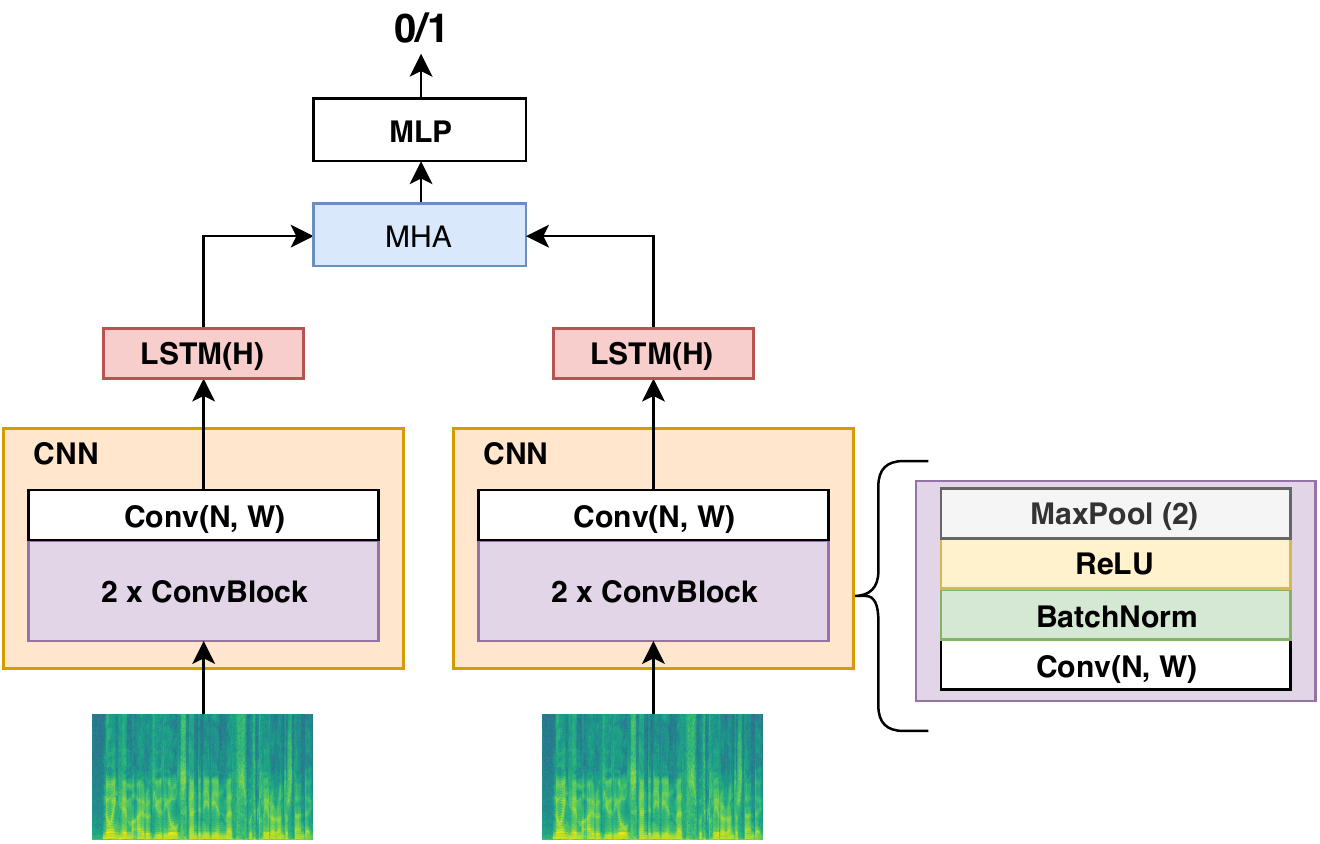}
    \caption{Architecture of the proposed siamese network. MHA: multi-head attention. MLP: multi-layer perceptron. CNN: convolutional neural network. Conv($N$, $W$) refers to a 1-dimensional convolutional neural layer with $N$ kernels and $W$ kernel length. For the LSTM block, ($H$) refers to having $H$ cells in the LSTM layer.}
    \label{fig:network}
\end{figure}

The cross-correlation method works upon raw features like the $\log$-magnitude spectrum, whereas deep learning approaches build intermediate abstractions of the features throughout a stack of neural layers~\cite{goodfellow2016deep}.\ These abstractions are usually more robust to low-level feature changes (like temporal shifts) and also facilitate a learnable preprocessing to solve a task of a high-abstraction level like classification.\ Hence we leverage the capacity of neural networks to process two separate $\log$-magnitude spectrum sources, the hypothesis and the reference, and to classify whether there is sample loss or not in the hypothesis.
To that end, we design a network that has two branches processing the raw spectral inputs, and relates them in the end with an attention mechanism.\ The proposed model is depicted in Figure~\ref{fig:network}.\ Given the hypothesis input and the reference input, both of length $\ell$, we inject them into the two encoder branches.\ Each input branch is composed by a convolutional neural network~(CNN) front-end, specially suitable to detect local temporal correlations in the spectral frames, and a long-short term memory~(LSTM)~\cite{hochreiter1997long} block that exploits specially the long-term sequential dependencies of the input sequence.\ The convolutional front-end also decimates the sequence lengths of the inputs by 4 through a couple of max-pooling layers that halve temporarily the feature maps right after each of the first two convolutional blocks.\ This type of model is known as siamese network~\cite{bromley1994signature}, where both branches are constrained to share the same weights as they process the same type of input data.\ The optimization problem in the siamese network then leads to learning two separate embeddings which have positive inner products for sequences of the same class but negative inner products for those of different classes~\cite{kim2019representation}.\ Once we obtained the two hidden representations of the signals, we forward them through the multi-head attention~(MHA) component, which outputs a weighted sum of the reference computed by a compatibility function of the hypothesis to itself~\cite{vaswani2017attention}.\ As a final step, we took the last time step values and forwarded it through a multi-layer perceptron (MLP), right before the application of a sigmoid function $\sigma$.

During training, the optimization procedure is not taking the multi-device scenario into account.\ The learning problem is reduced to the detection of the sample drop between pairs of hypotheses and references.\ Starting with the Equation \ref{eq:general} we define the function $\text{C}$ to be a composition between a multi-head attention block and a linear layer.\ By setting $\mathcal{F} = \sigma$, we have:
\begin{equation}
\label{eq:neural}
\sigma \left( \text{MLP} \left( \text{MHA}\left( h_{hyp}, h_{ref}  \right) \right)\right)   \rightarrow \{0,1\},
\end{equation}
 where $h_{hyp}$ and $h_{ref}$ are the embedding output by CNN-LSTM block.\ 
 Given a random variable $d$ with values in $\{0,1\}$ and indicating the sample drop event, the output distribution of the network can be defined as $p(d=1| h_{hyp}, h_{ref})$.\
 
 During the test time, for each device $j$ we forwarded five segment pairs using all devices $k \neq j$.\ The distinct output values of the network are then combined together by taking the average (or median) value, and leading to:
 \vspace{-0.2cm}
 \begin{equation}
     \label{eq:neural-test}
     \text {Avg}_k\left( \sigma \left(\text{MLP} \left(  \text{MHA}\left( h_{j}, h_{k}  \right) \right) \right)  \right)  \rightarrow \{0,1\}.
 \end{equation}

\section{Experiments}
\label{sec:exp}
The CHiME-5 corpus does not offer a reliable ground-truth for the sample drop detection task.\ Therefore, the overall training of the network has been carried out by using artificial data, and in two distinct stages.\ Concerning the evaluation of the performance of the model, we did not limit us to the mere usage of artificial data, and used also an evaluation set extracted from the real data.\

\subsection{Data}
\label{ssec:Ex-data}

\subsubsection{Noisy-reverberated LibriSpeech}
The first training round of the neural model was conducted using two different noisy and reverberated datasets, both derived from the LibriSpeech train-clean-100 corpus~\cite{panayotov2015librispeech}, containing 28539 speech utterances.\ Each utterance was filtered by using a different room Impulse Response (IR), characterized by a reverberation time typically ranging between 0.4 and 0.8 seconds.
IRs were computed by using a modified version of the image method~\cite{AllenBerkley,RavanelliOmologo}.\
Moreover, the two datasets differ both due to a different environmental noise, and to a different sound propagation time shift.\ 
Overall, SNRs are generally in the range between 5 and 25 dB.
Finally, a random sample loss event was simulated on one of the two versions, and from each of them segment pairs of hypothesis and reference were extracted.\

\subsubsection{Artificial multi-device mini-scenes}
The artificial dataset incorporates 1182 mini-scenes that were generated using clean Librispeech signals, and an accurate and realistic simulation of multi-microphone acquisition in a noisy and reverberant environment, as described in~\cite{RavanelliOmologo}.\ 
 Each scene refers to a different distribution of the six Kinect devices in space, with varying room size, location of speakers, and noise sources,
as well as directivity of source, microphone polar pattern, and reflection coefficient of each wall.\ The mini-scenes were created without any sample drops.\ Each scene was then postprocessed to simulate the introduction of none, one, or more drops, for a total number of 880 drops.\ The data set was split into 782, 100 and 300 mini-scenes for drop detection training, development and evaluation, respectively.\ The scene duration ranges from 5 to 30 seconds.\ This material has then been organized to expose the neural model to a balanced set of drop/no-drop examples.

\subsubsection{Real dataset}
In order to evaluate the proposed model in a more realistic context, it is necessary to test the performance also on a real dataset. In this regard, it is worth noting that there exists no real corpus for studies on sample drop detection. The evaluation set was created by extracting 65 segments, from which 31 with sample drop, from three sessions of the train portion of the CHiME-5 corpus, namely sessions 03, 07, and 08.\ 
Some of these segments are characterized by a loss that was found through a very careful visual inspection
of the related spectrograms, as outlined in Section \ref{ssec:classic}.\ 
\vspace{-0.2cm}
\subsection{Setting}
\label{ssec:Ex-set}
The duration of the loss for the two different contaminated versions of LibriSpeech, was drawn from a left-truncated normal distribution $\mathcal{N}(600,150)$, with a cut value of $50$. The samples belonging to that interval were then eliminated from one of the segments.\ A one second context window containing this loss was isolated, by positioning the loss point randomly within the context window.\ The parameters of the pre-trained model on this data were then used for a further training step on the mini-scenes.\ The performance of the final model was tested on two separate evaluation sets belonging to both artificial and real data.\
The labels of the one-second-long segment pairs are distributed equally in both training and evaluation sets.\
Regarding the network's architecture, two different experiments were carried out.\ First, we concatenated $h_{hyp}$ and $h_{ref}$ defined in Equation \ref{eq:neural} and applied the sigmoid function on the last time step.\ Then we added the attention block and operated in the same way for the binary output.\ We also experimented with the average activations over the time values, but it proved to be less effective. We trained the models for $20$ epochs on minibatches of size $50$ and $30$ for the pre-trained and final models respectively.\ We used the Adam optimizer~\cite{kingma2014adam} with the default PyTorch~\cite{paszke2017automatic} parameters and learning rate $5\cdot 10^{-5}$. Regarding the network structure, fully connected and convolutional layers comprise 512 units, with kernel length $N=5$ in the case of convolutions. The LSTM layer contains 1024 cells. The model structure hence amounts to approximately 15\,M of parameters in total.
\vspace{-0.2cm}
\subsection{Results}
\label{ssec:Ex-results}
The results obtained from different experiments are reported in Table \ref{tb:results}.\ 
A quick overview of the numbers reveals the importance of having a two-stage training procedure, since the pre-trained model on the contaminated LibriSpeech or the model trained directly on the mini-scenes without a pre-training phase have both a poor performance.\ The addition of the attention mechanism is also another important aspect.\ The difference of F1 score of the final model over mini-scenes is much smaller than the one over CHiME-5, in case of application of multi-head attention.\ Furthermore, the big gap between the result of the pre-trained model and others indicates the importance of having an appropriate data for multi-device case.\

For the Fast Fourier Transform (FFT) analysis frame length we tried 32 and 64 milliseconds~(ms) with 50\% overlap.\ The former choice leads to better results, specially in case of real data.\
During the test time we combined the output of the network for different pair devices by using major voting, averaging and the median value.\ The best choice resulted to be averaging and median for mini-scenes and CHiME-5, respectively.\ 


\vspace{-0.2cm}
\renewcommand{\arraystretch}{1.15}

\begin{table}[!ht]

	\setlength{\tabcolsep}{0.2em}  
	\centering
	\caption{\small{Precision~(P), Recall~(R) and F1 Score [\%] for the neural model pre-trained only on contaminated LibriSpeech (Pre-NN), and the final neural model (NN) with and without attention and pre-training step, on artificial multi-device mini-scenes and CHiME-5 corpus, using FFT analysis frame length of 32 ms.}}
	\vspace{1mm}
	
	\label{tb:results}
				\begin{tabular}{ |c|c|c|c|c|c|c|c|c|} \hline

			\multirow{2}{*}{Model} &\multirow{1}{*}{Pre-}&\multirow{2}{*}{Attention}& \multicolumn{3}{|c|}{Mini-scenes} & \multicolumn{3}{|c|}{CHiME-5}    \\ \cline{4-9}

			&Trained&& \scriptsize{P} & \scriptsize{R} & \scriptsize{F1}  &\scriptsize{P} & \scriptsize{R} & \scriptsize{F1}  \\ \hline

			Pre-NN &- &yes &  57.1 & 64.5 & 60.6 & 45.6 & 67.7 & 54.5 \\ \hline
				\multirow{3}{*}{NN}& no &yes & 48.6 & 81.6  & 60.9 & 48.0 & 93.0 & 63.3\\ \cline{2-9}
			
			 &\multirow{2}{*}{yes}& no  &77.8 &86.7 & 82.0 & 50.0 &74.1 &59.7 \\ \cline{4-9}
			 && yes   & 87.5 &  88.6   & \textbf{88.0}  &90.0  & 87.0  & \textbf{88.5}\\ \hline

		\end{tabular}
		
	\end{table}

\vspace{-0.2cm}

\section{Conclusions}
\label{sec:conclusion}
\vspace{-0.1cm}

In this paper, we investigated a possible approach for the detection of sample drops in the context of multi-microphone devices distributed in space, a very challenging technical issue that recently interested many researchers working on the CHiME-5 challenge.

The proposed approach consists in combining a normalized cross-correlation processing and a neural classifier, in order to detect short context windows that are characterized by a possible loss.\ Experimental results show that a classification performance of 88\% F1 score is obtained both on simulated and on real evaluation data sets.\ In the near future, we plan to improve this performance along different research directions, such as cross-processing all the microphone signals acquired by the available devices.

Though our current main focus is on the CHiME-5 data set the proposed solution can be applied to other similar contexts, which are affected by loss of segments in the audio input signals.
Our work also represents a preliminary step towards a joint combination of sample drop detection and quantification of the duration of the lost segment, which is a research issue under study.\ Moreover, we envisage a third processing step that includes a possible reconstruction of the lost information, based on exploiting redundant information available from higher-quality input channels not affected by the sample drop.
For all of these foreseen directions, we also plan to verify soon a possible improvement in terms of recognition performance on the CHiME-5 task provided by the application of the proposed method and of the resulting array synchronization.

\vspace{-0.2cm}
\section{Acknowledgements}
\vspace{-0.1cm}

\label{sec:acknowledgements}
The work reported here was started at JSALT 2019, and supported by JHU with gifts from Amazon, Facebook, Google, and Microsoft. This work was also supported by the project TEC2015-69266-P (MINECO/FEDER, UE).

We thank Jon Barker for providing us with a preliminary list of temporal instants related to possible sample drops in CHiME-5.

We would like also to express our gratitude to Tobias Menne and Ralf Schl\"uter for their valuable feedback and suggestions.

\vfill\pagebreak

\bibliographystyle{Sample-drop}
\bibliography{strings,refs}

\end{document}